\documentclass[aps,showpacs,floatfix,twocolumn,superscriptaddress,prl]{revtex4}

\usepackage{epsfig,amsmath,graphicx,amssymb,float,,bm}

\begin{document}
\title{Optimum spin-squeezing in Bose-Einstein condensates with particle losses}

\author{Yun Li}
\affiliation{Laboratoire Kastler Brossel, ENS, UPMC,
 24 rue Lhomond, 75231 Paris Cedex 05, France}

\affiliation{Department of Physics, East China Normal University,
Shanghai 200062, China}

\author{Y. Castin}
\affiliation{Laboratoire Kastler Brossel, ENS, UPMC,
 24 rue Lhomond, 75231 Paris Cedex 05, France}

\author{A. Sinatra}
\affiliation{Laboratoire Kastler Brossel, ENS, UPMC,
 24 rue Lhomond, 75231 Paris Cedex 05, France}

\begin{abstract}

The problem of spin squeezing with a bimodal condensate in presence of particle losses 
is solved analytically by the Monte Carlo wavefunction method. 
We find the largest obtainable spin squeezing
as a function of the one-body loss rate, the two-body and three-body
rate constants, and the $s$-wave scattering length.

\end{abstract}
\pacs{}

\maketitle

Spin squeezed states, first introduced in \cite{Ueda}, generalize to
spin operators the idea of squeezing developed in quantum optics. In
atomic systems effective spins are collective variables that can be
defined in terms of two different internal states of the atoms
\cite{expPolzik} or two orthogonal bosonic modes \cite{Sorensen}.
States with a large coherence between the two modes, that is with a
large mean value of the spin component in the equatorial plane of
the Bloch sphere, can still differ by their spin fluctuations. For
an uncorrelated ensemble of atoms, the quantum noise is evenly
distributed among the spin components orthogonal to the mean spin.
However quantum correlations can redistribute this noise and reduce
the variance of one spin quadrature with respect to the uncorrelated
case, achieving spin squeezing. Besides applications in quantum
communication and quantum information \cite{Molmer}, these
multi-particle entangled states have practical interest in atom
interferometry, and high precision spectroscopy \cite{Wineland}
where they could be used to beat the standard quantum limit already
reached in atomic clocks \cite{Santarelli}.

Different techniques to create spin squeezed states in atomic
systems have been proposed and successfully realized experimentally
including transfer of squeezing from light to matter
\cite{LighttoAtoms} and quantum non demolition measurements of the
atomic state \cite{QND}. To go further, it was shown that coherent
interactions between cold atoms in a bimodal Bose-Einstein
condensates \cite{Sorensen} can in principle provide a huge amount
of entanglement and spin squeezing. It is thus important to
determine the ultimate limitations imposed by decoherence to the
maximum spin squeezing that can be obtained by this method. Several
forms of decoherence may be present in the experiment. 
The case of a dephasing perturbation was studied in \cite{Lukin_deph}.
In this work we deal with particle losses, an unavoidable source of
decoherence in cold atom systems, due e.g. to collisions of
condensed atoms with the hot background gas, and to three-body
collisions leading to molecule formation.

As shown in \cite{Sorensen}, bimodal Bose-Einstein condensates
realize the one-axis twisting model proposed in \cite{Ueda} to
create spin squeezing. This exactly solvable model predicts a perfect
squeezing in the limit of a very large system: Formally $\xi^2 \to 0$
for $N \to \infty$, $N$ being the number of particles in the system 
and $\xi^2$ the squeezing parameter defined in Eq.(\ref{eq:defxi2}).
We expect losses to degrade the squeezing \cite{Sorensen} 
that is $\xi^2_{\rm no\ loss} \leq \xi^2_{\rm with\ loss}$ for any value of $N$.
However, as $\xi^2_{\rm no\ loss} \to 0$ as $N\to \infty$, 
this inequality does not tell us what will be the best squeezing 
in presence of losses.
In particular the limit $\,\lim_{N\to +\infty} \xi^2_{\rm with\ loss}$ 
could be zero
(perfect squeezing), a very small constant, or a constant close to 
one (one meaning no squeezing). 
We show that the second possibility is the correct one. 
The best achievable squeezing is reached for $N\to+\infty$,
and we derive its explicit expression,
as a function of the scattering length
and the loss constants $K_1,K_2,K_3$. 

%%%%%%%%%%%%%%%%%%%%%%%%%%%%%%%%%%%%%%%%%%%%%%%%%%%%%%%%
We consider two spatially separated symmetric condensates $a$ and
$b$ prepared in an initial state with $N$ particles and a well
defined relative phase \cite{noteAlice}
\begin{equation}
|\phi \rangle \equiv \frac{1}{\sqrt{N!}}\left( \frac{e^{i\phi}
a^\dag+e^{-i\phi} b^\dag} {\sqrt{2}}\right)^N|0\rangle \,.
\label{eq:initial}
\end{equation}
We assume that $\phi=0$ initially. Correspondingly, the $x$
component of the collective spin $S_x=(a^\dagger b + b^\dagger a)/2$
has a mean value $\langle S_x \rangle=N/2$. Here we assume that no
excitation is created during the preparation process and we neglect
all the other modes than the condensate modes $a$ and $b$. When
expanded over Fock states $|N_a,N_b\rangle$, the state
(\ref{eq:initial}) shows binomial coefficients which, for large $N$,
are peaked around the average number of particles in $a$ and $b$,
$\bar{N}_a=\bar{N}_b=N/2$. We use this fact to approximate the
Hamiltonian with its quadratic expansion around $\bar{N}_a$ and $\bar{N}_b$ \cite{Sinatra}: 
%\begin{equation}
$
H_0=\sum_{\epsilon=a,b} E(\bar{N}_\epsilon)+\mu_\epsilon(\hat{N}_\epsilon-\bar{N}_\epsilon)
+ \frac{1}{2} \mu_\epsilon^\prime (\hat{N}_\epsilon-\bar{N}_\epsilon)^2 
$
%\end{equation}
where $\mu_\epsilon$ is the chemical potential for the $\epsilon$ condensate and
$\mu_\epsilon^\prime \equiv (\partial_{N_\epsilon}\mu_\epsilon)_{\bar{N}_\epsilon}$.
In the symmetric case, we can write
\begin{equation}
H_0=f(a^\dag a + b^\dag b) + \frac{\hbar \chi}{4} (a^\dag a -b^\dag b)^2 
\end{equation}
where $\chi=\mu_a^\prime/\hbar$.  The first term in $H_0$ is some function $f$
of the total atom number: It commutes with the 
density operator $\rho$ of the system and can be omitted.

In presence of one, two and three-body losses, the evolution of the density
operator, in the interaction picture with respect to $H_0$,
is ruled by the master equation 
\begin{equation}
\frac{d\tilde{\rho}}{dt}=\sum_{m=1}^{3} \sum_{\epsilon=a,b}
\gamma^{(m)}\left[{c}^m_\epsilon\tilde{\rho}{c}^{\dag
m}_\epsilon-\frac{1}{2}\{{c}^{\dag m}_\epsilon{c}^m_ \epsilon,
\tilde{\rho}\} \right]
\end{equation}
where $\tilde{\rho}=e^{iH_0t/\hbar} \rho e^{-iH_0t/\hbar}$,
${c}_a=e^{iH_0t/\hbar} a e^{-iH_0t/\hbar}$, and similarly for $b$,
$\gamma^{(m)}=\frac{K_m}{m}\int d^3r|\phi(r)|^{2m}$, where $K_m$ is
the $m$-body rate constant and $\phi(r)$ is the condensate
wavefunction in one of the two modes. In the Monte Carlo
wavefunction approach \cite{MCD} we define an effective Hamiltonian
$H_{\text{eff}}$ and the jump operators $J^{(m)}_\epsilon$
\begin{eqnarray}
H_{\text{eff}}&\hspace{-8mm}=&\hspace{-7mm} - \sum_{m=1}^{3}
\sum_{\epsilon=a,b} \frac{i\hbar}{2}\gamma^{(m)} c_
\epsilon^{\dag m}c_\epsilon^m  \,; \label{H_eff} \\
&\hspace{1mm}J^{(m)}_\epsilon&\hspace{-1mm}=\sqrt{
\gamma^{(m)}}{c}^m_\epsilon\,.
\end{eqnarray}
We assume that a small fraction of particles will be lost during the
evolution so that we can consider $\chi$ and $\gamma^{(m)}$
($m=2,3$) as constant parameters of the model. The state evolution
in a single quantum trajectory is a sequence of random quantum jumps
at times $t_j$ and non-unitary Hamiltonian evolutions of duration
$\tau_j$:
\begin{eqnarray}
|\psi(t)\rangle&\hspace{-2mm}=&\hspace{-1mm}e^{-iH_{\text{eff}}
(t-t_k)/\hbar}J^{(m_k)}_{ \epsilon_k}(t_k)e^{-iH_{\text{eff}}
\tau_k/\hbar}J^{(m_{k-1})}_{
\epsilon_{k-1}}(t_{k-1}) \nonumber \\
&&\ldots J^{(m_1)}_{\epsilon_1}(t_1)e^{-iH_{\text{eff}}
\tau_1/\hbar}|\psi(0)\rangle\,. \label{eq:evol}
\end{eqnarray}
The expectation value of any observable $\hat{\mathcal{O}}$ is
obtained by averaging over all possible stochastic realizations,
that is all kinds, times and number of quantum jumps, each
trajectory being weighted by its probability \cite{MCD}
\begin{equation}
\langle\hat{\mathcal {O}}\rangle=\sum_{k}\int_{0<t_1<t_2<\cdots
t_k<t}\hspace{-2cm} dt_1dt_2\cdots dt_k \sum_{\{\epsilon_j, m_j\}}
\langle\psi(t)|\hat{\mathcal {O}}|\psi(t)\rangle\,.
\end{equation}
We want to calculate spin squeezing. In the considered symmetric
case with zero initial relative phase, the mean spin remains aligned
to the $x$ axis $\langle S_x \rangle=\langle b^\dag a \rangle$, and
the spin squeezing is quantified by the parameter
\cite{Wineland,Sorensen}
\begin{equation}
\xi^2=\min_\theta \frac{\langle\hat{N}\rangle\Delta
S^2_\theta}{\langle S_x \rangle^2} \,, \label{eq:defxi2}
\end{equation}
where $S_\theta= (\cos \theta) S_y +(\sin \theta) S_z$,
$S_y=(a^\dagger b-b^\dagger a)/(2i)$, $S_z=(a^\dagger a-b^\dagger
b)/2$ and $\hat{N}=a^\dagger a + b^\dagger b$. The non correlated
limit yields $\xi^2=1$, while $\xi^2<1$ is the mark of an entangled
state \cite{Molmer,Sorensen}. In all our analytic treatments, it
turns out that $\Delta S_z^2=\langle \hat{N} \rangle/4$. This allows
to express $\xi^2$ in a simple way:
\begin{equation}
\xi^2=\frac{\langle a^\dagger a\rangle}{\langle b^\dagger a
\rangle^2}\left( \langle a^\dagger a\rangle + {A} -
\sqrt{{A}^2+{B}^2}\right)\,, \label{eq:def2xi2}
\end{equation}
with
\begin{eqnarray}
A &=& \frac{1}{2}\, \mbox{Re}\left( \langle b^\dagger a^\dagger a b - b^\dagger b^\dagger a a\rangle \right) \\
\label{eq:A} B &=& 2 \; \mbox{Im} \left( \langle b^\dagger b^\dagger
b a \rangle \right) \,. \label{eq:B}
\end{eqnarray}
With one-body losses only, the problem is exactly solvable.
Following a similar procedure as in \cite{Sinatra}, we get
\begin{equation}\label{xi_1b}
\xi^2(t)=\frac{ 1+\frac{1}{4}(N-1)e^{-\gamma
t}[\tilde{A}-\sqrt{\tilde{A}^2+\tilde{B}^2}] } {
\left[\dfrac{\gamma^2+\chi [\gamma \sin (\chi t) + \chi \cos (\chi
t)]e^{-\gamma t}} {\gamma^{2}+\chi^2} \right]^{2N-2}  }
\end{equation}
with $\gamma\equiv \gamma^{(1)}$ and
\begin{eqnarray*}
&\hspace{-2mm}\tilde{A}&\hspace{-2mm}=\hspace{-1mm}1-\left[\frac{\gamma^2+2\chi
[\gamma \sin (2\chi t) + 2 \chi \cos (2 \chi t)]e^{-\gamma t}}
{\gamma^2+4\chi^2} \right]^{N-2} \\
&\hspace{-2mm}\tilde{B}&\hspace{-2mm}=\hspace{-1mm}4 \sin \chi t
\left[\frac{\gamma^2+ \chi [\gamma \sin (\chi t) + \chi \cos (\chi
t) ]e^{-\gamma t}} {\gamma^{2}+\chi^2} \right]^{N-2} .
\end{eqnarray*}
The key points are that (i) $H_{\text{eff}}$ is proportional to
$\hat{N}$ so it does not affect the state, and (ii) a phase state
$|\phi\rangle$ is changed into a phase state with one particle less
after a quantum jump, $c_{a,b}|\phi\rangle\propto |\phi \mp \chi
t/2\rangle$ where $t$ is the time of the jump, the relative phase
between the two modes simply picking up a random shift $\mp\chi t/2$
which reduces the squeezing.

\begin{figure}
\includegraphics[width=8cm]{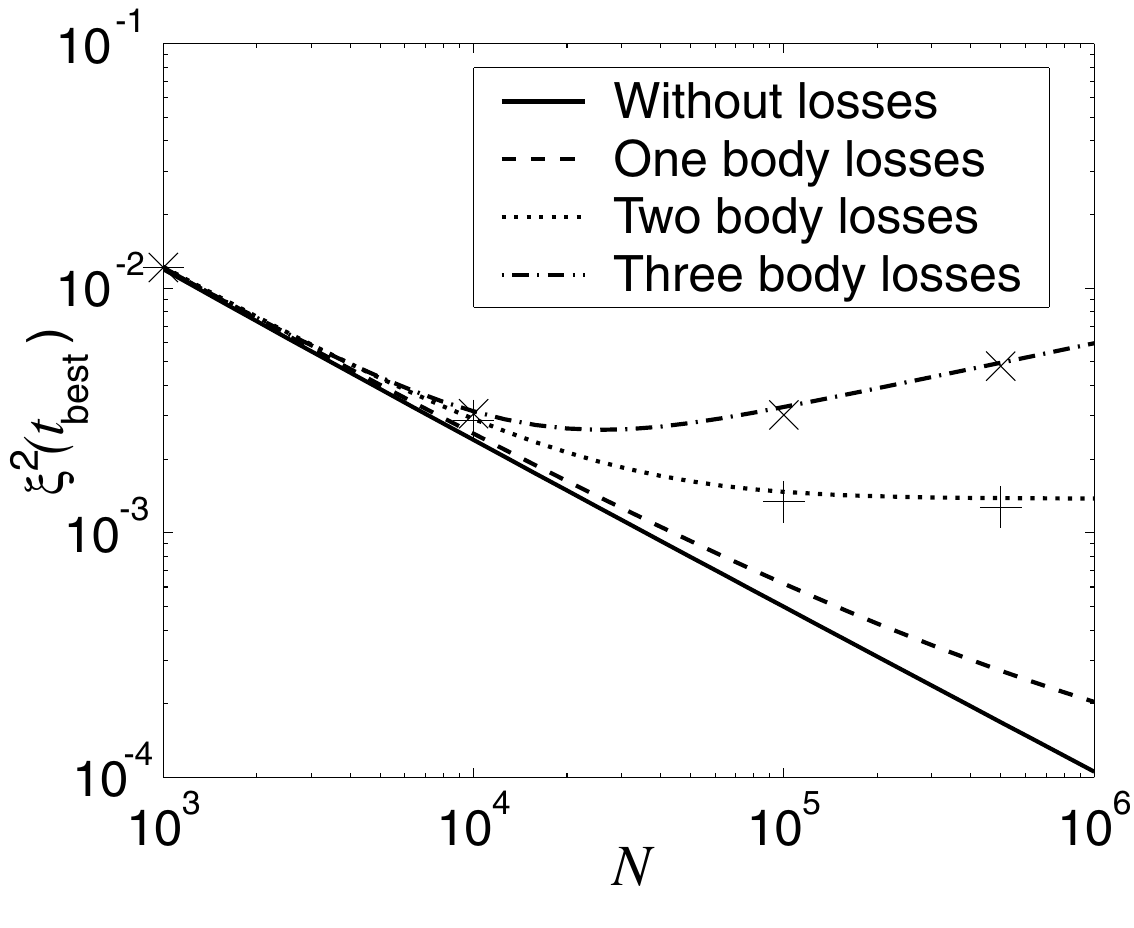}
\caption{Spin squeezing obtained by a minimization of $\xi^2$ over
time, as a function of the initial number of particles, without loss
of particles (solid line), with one-body losses (dashed line), with
two-body losses (dotted line), with three-body losses (dash-dotted
line) respectively. Parameters: $a=5.32$nm, $\bar{\omega}=2\pi\times
200$Hz, $K_1=0.1$s$^{-1}$, $K_2=2\times10^{-21}$m$^3/$s
\cite{BMV}, $K_3=18\times 10^{-42}$m$^6/$s. 
The symbols: crosses (plus) are results of
a full numerical simulation with 400 Monte Carlo realizations for
two-body (three-body) losses.}
\label{Fig:1}
\end{figure}

When two and three-body losses are taken into account, an analytical
result can still be obtained by using a constant loss rate
approximation \cite{Sinatra}
\begin{equation}\label{H_eff_app}
H_{\text{eff}}\simeq  - \sum_{m=1}^{3} \sum_{\epsilon=a,b}
 \frac{i\hbar}{2} \gamma^{(m)} \bar{N}_\epsilon^{m} \equiv - \frac{i \hbar}{2} \lambda \,.
\end{equation}
We verified by simulation (see Fig.\ref{Fig:1}) that this is valid
for the regime we consider, where a small fraction of particles is
lost at the time at which the best squeezing is achieved. In this
approximation, the mean number of particles at time $t$ is
\begin{equation}
\langle\hat{N}\rangle=N[\,1-\sum_m {\Gamma}^{(m)}t\,]\,; \:\:\:
{\Gamma}^{(m)}\equiv(N/2)^{m-1}m\gamma^{(m)}
\end{equation}
where ${\Gamma}^{(m)}t$ is the fraction of lost particles due to
$m$-body losses. Spin squeezing is calculated from
(\ref{eq:def2xi2}) with
\begin{eqnarray}
\label{eq:ab} \langle b^\dagger a \rangle & =& \frac{e^{-\lambda
t}}{2} \cos^{N-1}(\chi t)
 \tilde{N} F_1
\\
\label{eq:A2} A &=& \frac{e^{-\lambda t}}{8} \tilde{N}(\tilde{N}-1)
\left[ F_0 - F_2 \cos^{N-2}(2\chi t)\right] \\
\label{eq:B2} B &=& \frac{e^{-\lambda t}}{2} \cos^{N-2}(\chi t)
\sin(\chi t) \tilde{N}(\tilde{N}-1) F_1
\end{eqnarray}
where the operator $\tilde{N}=(N-\partial_\alpha)$ acts on the
functions
\begin{equation}
F_\beta(\alpha)=\exp\left[ \sum_{m=1}^3 2 \gamma^{(m)}t e^{\alpha
m}\frac{\sin (m \beta \chi t)}{m\beta \chi t \cos^m(\beta \chi
t)}\right] \,,
\end{equation}
and all expressions should be evaluated in $\alpha=\ln \bar{N}_a$.

We want to find simple results for the best squeezing and the best
squeezing time in the large $N$ limit. In the absence of losses
\cite{Ueda} the best squeezing and the best squeezing time in units
of $1/\chi$ scale as $N^{-2/3}$. We then set $N=\varepsilon ^{-3}$
and rescale the time as $\chi t=\tau \varepsilon^2$. We expand the
results (\ref{xi_1b}) and (\ref{eq:ab}-\ref{eq:B2}) for $\varepsilon
\ll 1$ keeping ${\Gamma}^{(m)}/\chi$ constant, and we obtain in both
cases
\begin{equation}
\xi^2(t) \simeq \frac{1}{N^2(\chi t)^2}+\frac{1}{6}N^2(\chi
t)^4+\frac{1}{3}\Gamma_{\text{sq}}t\,,
\end{equation}
with
\begin{equation}
\Gamma_{\text{sq}}=\sum_m \Gamma_{\text{sq}}^{(m)} \ \ \ \mbox{and}
\ \ \ \Gamma_{\text{sq}}^{(m)} = m {\Gamma}^{(m)}\,.
\end{equation}
For equal loss rates ${\Gamma}^{(m)}$, the larger $m$, the more the
squeezing is affected. Introducing the squeezing $\xi^2_0(t)$ in the
no-loss case, the above result can be written as
\begin{equation}
\xi^2(t)=\xi^2_0(t)\left[1+\frac{1}{3} \frac{\Gamma_{\text{sq}}t}{
\xi^2_0(t)}\right].
\end{equation}
This shows that (i) the fact that only a small fraction of atoms is
lost at the best squeezing time does not imply that the correction
on the squeezing due to losses is small; (ii) the more squeezed the
state is, the more sensitive to the losses. 
%This was expected as a
%larger amount of correlation is then lost in each loss event. 
In presence of losses, the best squeezing time and the corresponding
squeezing are
\begin{eqnarray}
&&\hspace{2mm}t_{\text{best}}=\left[\frac{f(C)}{2}\right]^{1/3}
\frac{N^{-2/3}}{\chi}\,,\label{tbest}\\
\hspace{-3mm}\xi^2(t_{\text{best}})&\hspace{-2mm}=&\hspace{-2mm}\left[
\frac{1}{f(C)^{2/3}}+ \frac{f(C)^{4/3}}{24} +\frac{C
f(C)^{1/3}}{3}\right]
\hspace{-1mm}\left(\frac{2}{N}\right)^{2/3}\label{xi_tbest}
\end{eqnarray}
\vspace{-3mm}
\begin{equation}
f(C)=\sqrt{C^2 +12}-C \,; \hspace{2mm}
C=\frac{\Gamma_{\text{sq}}}{2\chi}.
\end{equation}
In order to find optimal conditions to produce spin squeezing in
presence of losses and set the ultimate limits of this technique,
from now on, we assume that the number of particles is large enough
for the condensates to be in the Thomas-Fermi regime so that
\begin{equation}
\mu=\frac{1}{2}\hbar\bar{\omega}\left(\frac{15}{2}
\frac{Na}{a_0}\right)^{2/5}\,,
\end{equation}
where $a_0=\sqrt{\hbar/M\bar{\omega}}$\, is the harmonic oscillator
length, $M$ is the mass of a particle and $\bar{\omega}$ is the
geometric mean of the trap frequencies,
\begin{eqnarray}
\hspace{-8mm}&&\chi=\frac{2^{3/5}3^{2/5}}{5^{3/5}}\left(\frac{\hbar}{M}
\right)^{-1/5}a^{2/5}\bar{\omega}^{6/5}N^{-3/5}\\
\hspace{-8mm}&& \Gamma^{(1)}=K_1\\
\hspace{-8mm}&& \Gamma^{(2)}=\frac{15^{2/5}}{2^{7/5}7\pi}
\left(\frac{\hbar}{M}\right)^{-6/5}a^{-3/5}
\bar{\omega}^{6/5}N^{2/5}K_2\\
\hspace{-8mm}& &\Gamma^{(3)}= \frac{5^{4/5}}{2^{19/5}3^{1/5}7 \pi^2}
\left(\frac{\hbar}{M}\right) ^{-12/5} a^{-6/5} \bar{ \omega} ^{12/5}
N^{4/5}K_3 \,.
\end{eqnarray}
We first analyze the dependence of squeezing on the initial number
of particles, separating for clarity one, two and three-body losses.
Fig.\ref{Fig:1} shows the best squeezing $\xi^2(t_{\text{best}})$ as
a function of $N$ when only one kind of losses is present.
The curve without losses is also shown for
comparison. According to Fig.\ref{Fig:1}, one-body losses do not
change qualitatively the picture without losses and we have
$\xi^2(t_{\text{best}}) \propto N^{-4/15}$ for $N\rightarrow
\infty$. In the same limit, with two-body losses,
$\xi^2(t_{\text{best}})$ is independent of $N$. With three-body
losses, $\xi^2(t_{\text{best}}) \propto N^{4/15}$ for $N\rightarrow
\infty$, implying that, for a fixed $\bar{\omega}$, there is a finite optimum number of particles
for squeezing.

%===========================================================%
\begin{figure}
\includegraphics[width=6.25cm]{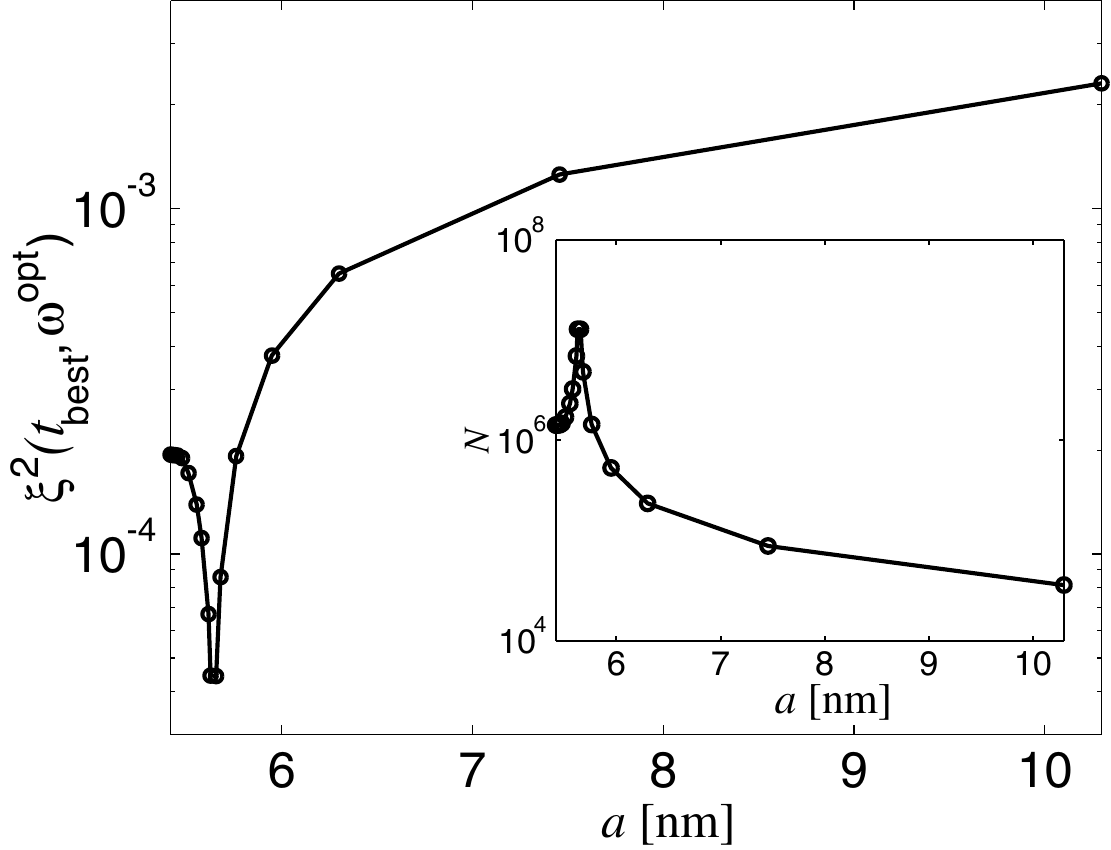}
\caption{Spin squeezing $\xi^2({t_{\text{best}}})$ optimized with
respect to $\bar{\omega}$ as a function of the scattering length
$a$, when the magnetic field is varied on the left side of the
$B_0=1007.4$G Feshbach resonance of ${}^{87}$Rb. The inset shows the
number of particles for each point, calculated for $\eta=10\%$. We
took $a(B)=a_{\text{bg}}[1-\Delta B/(B-B_0)]$ with
$a_{\text{bg}}=5.32$nm, $\Delta B=0.21$G. The three-body rate
constant $K_3(B)$ is taken from \cite{SGK}, $K_1=0.01$s$^{-1}$ and
$K_2=0$.}\label{Fig:2}
\end{figure}

%===========================================================%
We now turn to a full optimization of squeezing over $\bar{\omega}$
and $N$ in the simultaneous presence of one, two and three-body
losses. To this end, we note that the square brackets in
Eq.(\ref{xi_tbest}) is an increasing function of $C$, we can then
optimize $\xi^2(t_{ \text{best}})$ by minimizing $C$ with respect to
$\bar{\omega}$. Under the conditions $K_1\neq 0$ and $K_3\neq 0$,
the minimum of $C$, $C_{\min}$, is obtained for
$\Gamma_{\text{sq}}^{(3)}=\Gamma_{\text{sq}}^{(1)}$ yielding
\begin{equation}
\bar{\omega}^{\text{opt}}= \frac{2^{19/12}7^{5/12}\pi^{5/6}}
{15^{1/3}} \frac{\hbar}{M}\frac{a^{1/2}}{N^{1/3}}
\left(\frac{K_1}{K_3}\right)^{5/12}. \label{eq:omopt}
\end{equation}
It turns out that $C_{\min}$ is proportional to $N$ and
$\xi^2(t_{\text{best}},\bar{\omega}^{\text{opt}})$ is a decreasing
function of $N$. The lower bound for $\xi^2$, reached for $N=\infty$ is then
\begin{equation}
\min_{t,\bar{\omega},N} \xi^2 = \left( \frac{5\sqrt{3}}{28\pi}
\frac{M}{\hbar a}\right)^{2/3}\hspace{-1mm}\left[\sqrt{\frac{7}{2}
(K_1 K_3)}+K_2\right]^{2/3}. \label{eq:lboundxi2}
\end{equation}
In practice, one can choose $N=N_\eta$ in order to have
$\xi^2=(1+\eta)\min \xi^2$ (e.g. $\eta=10\%$), and then calculate
the corresponding optimized frequency $\bar{\omega}^{\text{opt}}$
with (\ref{eq:omopt}).
For a suitable choice of the internal state, in an optical trap, the
two-body losses can be neglected $K_2=0$. One can get in this case
very simple formulas for the optimized parameters and squeezing.
For $\eta=10\%$ \cite{validity}:
\begin{eqnarray}
&&N_{\eta}\simeq\frac{17.833}{(K_1 K_3)^{1/2}}\frac{\hbar a}{M} \,,
\label{eq:N13}\\
&&t_{\text{best}}\simeq0.277\left(\frac{M}{\hbar K_1 }\right)^{2/3}
\left(\frac{K_3}{a^2}\right)^{1/3} ,
\label{eq:tbest13}\\
&&\xi^2\simeq0.356\left(\frac{M K_1}{\hbar}\right)^{1/3}
\left(\frac{M K_3}{\hbar a^2}\right)^{1/3}. \label{eq:xi13}
\end{eqnarray}
We now ask whether
%, in presence of one and three-body losses only,
we can use a Feshbach resonance to change the scattering length (but
also $K_3$) to improve the squeezing. 
In Fig.\ref{Fig:2} we plot the squeezing parameter {\sl
vs} the scattering length $a$. Predicted values of $K_3$, as a
function of $a$, are taken from \cite{SGK} for $^{87}$Rb in the
state $|F=1,m_F=1\rangle$ and $K_1=0.01$s$^{-1}$. We calculate
$\bar{\omega}^{\text{opt}}$ and the number of particles needed for
$\eta=10\%$ for each point in the curve. The dip giving large
squeezing corresponds to a strong decrease in $K_3$ around $1003.5$G
($K_3\simeq 3 \times 10^{-45}$m$^6/$s). Close to the Feshbach
resonance the squeezing gets worse as $K_3$ increases (even if in
the figure we do not enter the regime $K_3\sim \hbar a^4/m)$.

Finally we consider the problem of the survival time of a spin
squeezed state in presence of one-body losses. We imagine that the
system evolves in two periods: for $t<T_1$
the system is squeezed in presence of interactions
($\chi\neq 0$), one and three-body losses; and for $t>T_1$ 
the interaction is stopped ($\chi= 0$),
e.g.\ by opening the trap, and the system only experiences
one-body losses. As $t$ can be arbitrarily long, we
use the exact solution for $t>T_1$ 
%i.e. $H_{\text{eff}}$ given by  Eq.(\ref{H_eff}),
while for the $t<T_1 \simeq t_{\text{best}}$,
we use the approximation (\ref{H_eff_app}).
%i.e.  $H_{\text{eff}}$ given by Eq.(\ref{H_eff_app}).
Then for $t=T_1+T_2>T_1$:
\begin{eqnarray}\label{xi_2pd}
\xi^2(t)&\hspace{-2mm}=&\hspace{-1mm}\frac{1}{4} \frac{\langle
\hat{N}(T_1)\rangle^2}{\langle S_x(T_1)\rangle^2}-\left[
\frac{1}{4}\frac{\langle \hat{N}(T_1)\rangle^2}{\langle
S_x(T_1)\rangle^2}-\xi^2(T_1)\right]
e^{-\gamma^{(1)}T_2} \nonumber \\
&\hspace{-2mm}\simeq & \hspace{-1mm}
1-\left[1-\xi^2(T_1)\right]e^{-\gamma^{(1)}T_2}\,.
\end{eqnarray}
%In the second line of Eq.(\ref{xi_2pd}) we used the fact that for
%$T_1 \simeq t_{\text{best}}$, $\langle S_x\rangle\simeq \langle
%\hat{N}(T_1)\rangle/2$. 
This result shows that the spin squeezing
can be kept some time after the interactions have been stopped. To
give an example, for $^{87}$Rb atoms with bare scattering
length $a=5.32$nm, $K_1=0.01$s$^{-1}$, $K_2=0$, $K_3=6 \times
10^{-42}$m$^6/$s \cite{BGW}, in optimized conditions
(\ref{eq:N13})-(\ref{eq:xi13}) $N=2.8\times10^{5}$ and
$\bar{\omega}^ {\text{opt}}=2\pi \times20.06$Hz, $\xi^2=5.7\times
10^{-4}$ is reached at $T_1=t_{\text{best}}=4.4\times 10^{-2}$s, and
a large amount of squeezing $\xi^2\simeq 0.01$ is still available
after 1s.

%===========================================================%
%===========================================================%

In conclusion, we found the maximum spin-squeezing
reachable with cold atoms having a $S_z^2$ Hamiltonian, 
in presence of decoherence (losses)
unavoidably accompanying the elastic interaction among atoms. 
The best squeezing is reached for an atom number
$N\to \infty$ and not for a finite value of $N$.
This is important for applications such as spectroscopy where, 
apart from the gain due to
quantum correlations among particles (squeezing), 
one always gains in increasing $N$.

LKB is a unit of ENS and UPMC
associated to CNRS. We acknowledge discussions with the atom chip
team of Jakob Reichel. Y. Li acknowledges support from the ENS/ECNU
program.


\begin{thebibliography}{0}
\expandafter\ifx\csname natexlab\endcsname\relax\def\natexlab#1{#1}\fi
\expandafter\ifx\csname bibnamefont\endcsname\relax
  \def\bibnamefont#1{#1}\fi
\expandafter\ifx\csname bibfnamefont\endcsname\relax
  \def\bibfnamefont#1{#1}\fi
\expandafter\ifx\csname citenamefont\endcsname\relax
  \def\citenamefont#1{#1}\fi
\expandafter\ifx\csname url\endcsname\relax
  \def\url#1{\texttt{#1}}\fi
\expandafter\ifx\csname urlprefix\endcsname\relax\def\urlprefix{URL }\fi
\providecommand{\bibinfo}[2]{#2}
\providecommand{\eprint}[2][]{\url{#2}}

\end{thebibliography}


\begin{thebibliography}{99}

\bibitem{Ueda} M. Kitagawa and M. Ueda, Phys. Rev. A {\bf 47}, 5138 (1993).

\bibitem{expPolzik} B. Julsgaard, A. Kozhekin, E. Polzik, Nature {\bf 413}, 400 (2001).

\bibitem{Sorensen} A. S{\o}rensen, L.M. Duan, I. Cirac, P. Zoller,  Nature
{\bf 409}, 63 (2001).

\bibitem{Molmer} A. S{\o}rensen, K. M{\o}lmer, Phys. Rev. Lett.
{\bf 86}, 4431 (2001).

\bibitem{Wineland} D. J. Wineland, J.J. Bollinger,
W.M. Itano, D.J. Heinzen, Phys. Rev. A {\bf 50}, 67 (1994).

\bibitem{Santarelli} G. Santarelli \emph{et al}, Phys. Rev. Lett. {\bf
82}, 4619 (1999).

\bibitem{LighttoAtoms} J. Hald \emph{et al}., 
Phys. Rev. Lett. {\bf 83}, 1319 (1999);
% Expt. Transfer Squeezed Light --> squeezed atoms;
A. E. Kozhekin, K. M{\o}lmer, E. Polzik, Phys. Rev. A {\bf 62},
033809 (2000); 
A. Dantan, M. Pinard, Phys. Rev. A {\bf 69}, 043810 (2004).

\bibitem{QND} A. Kuzmich, L. Mandel, N. Bigelow, Phys. Rev. Lett.
{\bf 85}, 1594 (2000);
% Expt. Spin squeezing via QND measurement;
JM. Geremia, J. Stockton, H. Mabuchi, Science {\bf 304}, 270 (2004).
% Expt. Spin squeezing via QND measurement + feedback

\bibitem{Lukin_deph} A. M. Rey, L. Jiang, M. D. Lukin,
Phys. Rev. A {\bf 76}, 053617 (2007).

\bibitem{noteAlice}The case of two condensates in the same spatial mode and different
internal states, in the absence of demixing instability, see
\cite{Sorensen}, can be treated with minor modifications along the
same lines.

\bibitem{Sinatra} A. Sinatra, Y. Castin, Eur. Phys. J. D {\bf 4}, 247 (1998).

\bibitem{MCD} K. M{\o}lmer, Y. Castin, J. Dalibard, J. Opt. Soc. Am.
B {\bf 10}, 524 (1993); H. J. Carmichael, \emph{An Open Systems
Approach to Quantum Optics} (Springer
1993).


\bibitem{BMV} H. M. J. M. Boesten, A. J. Moerdijk, 
 B. J. Verhaar, Phys. Rev. A {\bf 54}, R29 (1996).

\bibitem{validity}
Our result is valid for $\min\, \xi^2 \ll 1$ as the fraction
of lost particles at time $t_{\rm best}$ is $\sim \min \xi^2$.

\bibitem{SGK} G. Smirne \emph{et al}, Phys. Rev. A {\bf 75}, 020702(R)
(2007).

\bibitem{BGW} E. A. Burt \emph{et al}, Phys. Rev. Lett. {\bf 79}, 337
(1997).

\end{thebibliography}
\end{document}